# Non-Equilibrated Counter Propagating Edge Modes in the Fractional Quantum Hall Regime

Anna Grivnin, Hiroyuki Inoue, Yuval Ronen, Yuval Baum, Moty Heiblum, Vladimir Umansky and Diana Mahalu

*Braun Center for Submicron Research, Department of Condensed Matter Physics,*

*Weizmann Institute of Science, Rehovot 76100, Israel*

**It is well established that 'density reconstruction' at the edge of a two dimensional electron gas takes place for 'hole-conjugate' states in the fractional quantum Hall effect (such as $v=2/3$, $3/5$, etc.). Such reconstruction leads, after equilibration between counter propagating edge channels, to a downstream chiral current edge mode accompanied by upstream chiral neutral modes (carrying energy without net charge). Short equilibration length prevented thus far observation of the counter propagating current channels – the hallmark of density reconstruction. Here, we provide evidence for such non-equilibrated counter-propagating current channels, in short regions ($l=4\mu m$ and $l=0.4\mu m$) of fractional filling $v=2/3$, and, unexpectedly, $v=1/3$, sandwiched between two regions of integer filling $v=1$. Rather than a two-terminal fractional conductance, the conductance exhibited a significant ascension towards unity quantum conductance ($G_Q=e^2/h$) at or near the fractional plateaus. We attribute this conductance rise to the presence of a non-equilibrated downstream $v=1$ channel in the fractional short regions.**

Ever since the realization that current in quantum Hall systems is carried by chiraly flowing electrons near the edge of the two dimensional electron gas (2DEG) [1,2], the study of edge dynamics has gained interest both theoretically and experimentally. While in the integer quantum Hall effect (IQHE) regime the picture of chiral edge channels is better understood, it is not the case in the fractional regime. It was suggested by Chang and Beenakker [3,4] that in the FQHE, very much like in the IQHE [5], the filling factor (density) decreases monotonically towards the edge of the electron gas, and the current is carried *downstream* by chiral channels in incompressible strips [3]. However, for hole-conjugate states, namely, $n+1/2<v<n+1$, with $n=0,1,2,…$ (say, $v=2/3$, $3/5$, etc.), MacDonald and Wen [6–8] proposed that the density profile near the edge is non-monotonic (named, *reconstructed*); thus supporting multiple propagating (*downstream*) or counter-propagating (*upstream*) edge channels. For example, based on the hierarchical picture [9,10], the $v=2/3$ state supports two counter-propagating chiral channels: a *downstream* electron channel, $v=1$, and an upstream $v=1/3$ channel - never observed [11]. Moreover, rather than a two-terminal conductance of the $v=2/3$ $G=4/3\times e^2/h$, the measured conductance should be $2/3\times e^2/h$. This, motivated Kane *et al.* [12,13], and later Meir *et al.* [14,15] to propose equilibration among the edge channels due to inter-channel Coulomb interaction accompanied by inter-channel tunneling due to disorder; leading to a downstream $v=2/3$ charged mode and an upstream neutral mode. The latter one was recently observed via shot noise measurements [16–18] and also by heating a narrow constriction [16,19–21]. Noting that for a sufficiently shallow confining potential near the edge, edge reconstruction was predicted to take place also in a variety of non-hole-conjugate fractional states [22–29] (neutral modes had been recently found in electron-like fractional states [18]) and even in integer fillings [30].

Our experiment was designed to study the $v=2/3$ and $v=1/3$ states before they reach equilibration, by varying the edge channels' propagation length in the short fractional regions; being in the range $l=0.4$-$4\mu m$. The 'fractional regions' (with filling $v_f$) were induced by top gates, each being sandwiched between two regions with bulk filling $v_B=1$ (Fig. 1). If a non-equilibrated edge channel in the fractional regions supports a downstream edge channel $v=1$, the transmission coefficient through it could ideally

(neglecting reflections) approach *unity*. However, if equilibration between the counter-propagating channels takes place in a fractional region, say, the $v_f=2/3$ region, the transmission through it would be $t=2/3$ – as is always observed in macroscopic samples [31,32].

Samples were fabricated on a variety of GaAs-AlGaAs heterostructures embedding high mobility 2DEG. A mesa was defined by wet etching and depleting metallic gates. Different widths fractional regions (45µm long denoted by LR and 0.1-4µm long denoted by SR) were defined by top gates (Fig. 1). An ingrown n$^+$:GaAs layer, 1µm below the 2DEG, served as a 'back gate'; allowing to vary the electron density in the range 0.5-2.2 $\times 10^{11}$cm$^{-2}$. The transmission through the fractional region was measured at different magnetic fields, different electron densities, and at different temperatures. Current was injected at the Source (S) with the voltage measured at the Probes (P1 and P2). Measurements were performed using a standard lock-in technique with excitation voltage ~2µV RMS at 179Hz at lattice temperature of 35mK (unless stated otherwise). Note that the excitation energy and temperature were substantially lower than the presumed energy gaps of $v=2/3, 1/3$ (~350µeV and higher) [33].

Our main results are summarized in Fig. 2a; showing the transmission (in units of the quantum conductance) from source S to probe P1 (through the 'sandwich' $v_B=1$ - $v_f$ - $v_B=1$) as function of the fractional region gate voltage (bulk density $n_B$~2.2×10$^{11}$cm$^{-2}$, $B$=8.5T). The transmission through LR is given, for reference, by the solid black line. Conductance plateaus at $v_f=2/3$ and $v_f=1/3$ are clearly visible, with somewhat weaker ones at $v_f=3/5$ and $v_f=2/5$ (and a signature of $v_f=2/7$). However, in structures with SR=4µm (dotted, blue) and 0.4µm (broken line, green) the conductance is not monotonic - exceeding the corresponding quantized values; rising markedly at the lower density side of the two main fractional plateaus (overlapping lower lying plateaus, with even a weak sign near the 2/5 plateau for SR=4µm). Note the large enhancement of the transmission for the SR=0.4µm, approaching 0.93×$e^2/h$ at $v_f=2/3$ – $v_f=3/5$, and 0.73×$e^2/h$ at $v_f=1/3$. Similar behavior was observed in other samples made from four different grown heterostructures with carrier densities in the range of (1.7 - 2.1)×10$^{11}$cm$^{-2}$, and with different SR lengths (see Supplementary Fig. S1). While the enhanced conductance at the hole-conjugate fractional fillings supports our hypothesis of a non-equilibrated edge channels at small enough propagation lengths, the observed enhancement in particle-like states is somewhat of a surprise – making the effect rather ubiquitous. However, it supports a proliferation of edge reconstruction points and presence of neutral modes, as indeed was recently observed [18].

Since upstream neutral modes in the integer regime were not observed, multiple non-equilibrated edge channels are not expected [34–39]. In these experiments, the bulk filling was set at $v_B$=2 ($B$=4.3T and density $2.2\times10^{11}$cm$^{-2}$) with the gated region tuned continuously from $v_f$ =2 to $v_f$ =0 (Fig 2b). The transmission through the LR shows multiple plateaus at (2, 5/3, 4/3, 1, 2/3, 1/3)$\times e^2/h$; however, the transmission through the SR was, as before, not monotonic. Conductance peaks are clearly observed near fractional fillings $v_f$=5/3, 2/3, and 1/3, but absent near $v_f$=1. At a lower magnetic field ($B$=2.8T and correspondingly lower density $1.4\times10^{11}$cm$^{-2}$), when no fractional states are formed, the conductance declined monotonically and showed nice plateaus of $2e^2/h$ and $e^2/h$ for all three sizes of the fractional regions (see Supplementary Fig. S2).

Edge reconstruction, being an outcome of electron interactions in a confining potential, is expected to depend on electron density and external magnetic field. We tested the effect of these variables in the three sizes of the fractional regime Fixing the density at $n_B$~$2.0\times10^{11}$cm$^{-2}$, transmission measurements were repeated at three different magnetic fields within the range of $v_B$=1 conductance plateau; $B$=8.5T, 7.7T, & 7.0T. The conductance is plotted for LR=45μm and SR=0.4μm Fig 3a (data for SR=4μm is shown in Supplementary Fig. S3a). While there was no visible effect in the LR sample, the transmission peak through the SR region tended to increase as the magnetic field was decreased. Next, the magnetic field was fixed at $B$=7T and the density was varied by charging the 'back-gate' in the range $n_B$=1.7 - $2.0\times10^{11}$cm$^{-2}$; again within the range of $v_B$=1 conductance plateau. Here, aside from the shift in the pinch off voltage, the conductance of LR=45μm and SR=0.4μm samples were effectively unaltered (see Fig. 3b; for data of SR=4μm see Supplementary Fig. S3b).

Temperature dependence was measured in order to rule out tunneling or activated transport through the narrow SR. With increasing temperature, the non-monotonic conductance peak near the fractional plateaus slightly diminished (Fig. 4a); suggesting an enhanced inter-channel tunneling and thus suppressing the equilibration length [26,40]. On the other hand, the transmission through the SR=0.4μm in the integer regime ($B$=2.8T) was found to increase steadily, starting already at $T$=230mK (Fig 4b). The conductance at 4K is plotted for reference in dash-dotted line (for data of SR=4μm see Supplementary Figs. S5 and S6).

Can the observed non-monotonic dependence of the transmission through the fractional regime be explained by a classical estimate of bulk currents in the Hall regime? In order to clarify that possibility we described the system by a standard 'Hall matrix' in each of the three regions; namely, $v_B$=1 - $v_f$ <1 - $v_B$=1, and a small longitudinal conductivity in each region. Solving the Poisson equation with the appropriate boundary conditions, we calculated the transmitted currents across the SR region as a function of its (classical) filling factor. The solution of the classical equations for SR=4μm coincides with the behavior of the conductance in the integer regime at $T$~4K; namely, when the QHE plateaus vanish (inset in Fig. 4b), justifying this approximation. When we compare the measured results in the fractional regime, say, at $v_f$=1/3 and at $v_f$=2/3, the conductance peaks are well *above* the calculated conductance (for more details, see Supplementary section *IV*). Though the numerical analysis leads to the same conclusion for SR<1μm, it is highly sensitive to changes in the spatial discretization scheme, and thus less reliable.

In a very recent work, we reported on the proliferation of neutral modes in many fractional states, upstream at the edge and throughout the incompressible bulk [18]; making edge reconstruction rather prevalent. Since neutral modes result after equilibration among multiple channels takes place, the nature of the non-equilibrated channels is generally concealed. Our observation, of an increased transmission of integer channels through fractional ones - described here in some detail for $v_f$=2/3 & 1/3, and also found in $v_f$=2/5, 4/5, and 5/3 - suggests edge reconstruction at these fractions, with an increase carrier density towards a filling of unity near the edge of the sample. Understanding details of edge reconstruction may allow manipulating it by innovative fabrication process. Since it is now clear that neutral modes persist throughout the fractional quantum Hall regime; hence likely to inhibit interference of fractional charges [41], their weakening (elimination) is desirable for the study anyonic statistics in the fractional regime [42].

We thank Ady Stern, Yuval Gefen and Yigal Meir for useful discussions. We acknowledge the partial support of the Israeli Science Foundation (ISF), the Minerva foundation, the U.S.-Israel Bi-National Science Foundation (BSF), the European Research Council under the European Community's Seventh Framework Program (FP7/2007-2013)/ERC Grant agreement No. 227716 and the German Israeli Project Cooperation (DIP).

# Figures

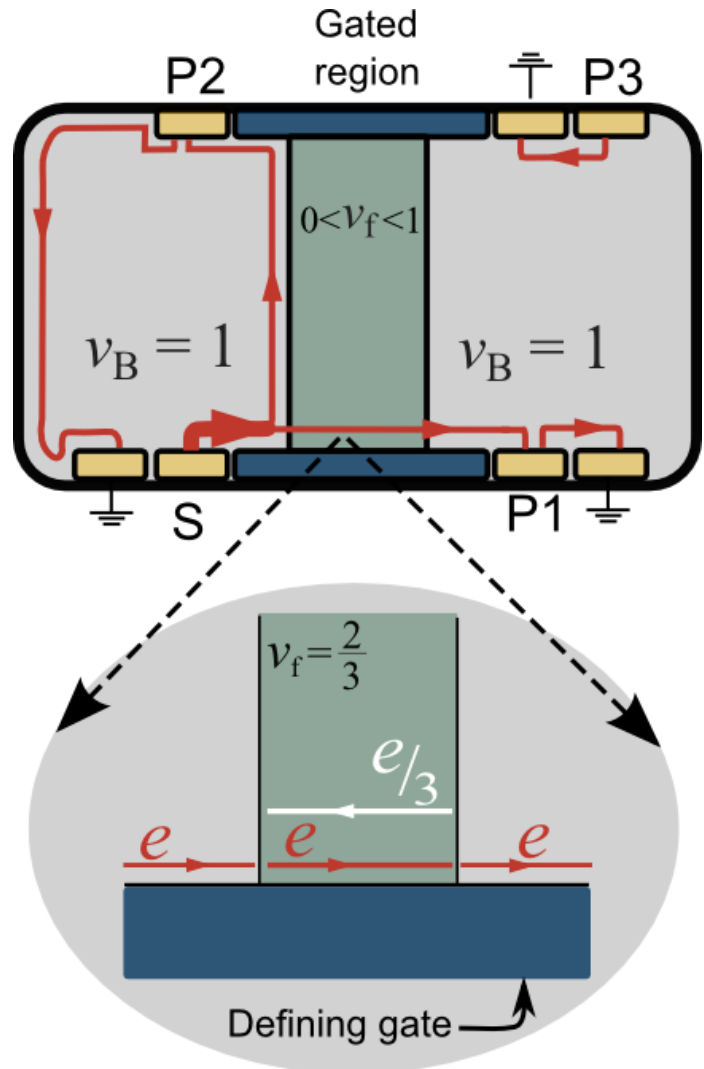

**FIG. 1.** Samples are fabricated in heterostructures embedding 2DEG with density $1.5\times10^{11}\text{cm}^{-2}$ with mobility $6\times10^{6}\text{cm}^{2}\text{V}^{-1}\text{s}^{-1}$ buried 85nm below the surface. Current is injected from the source contact S, towards a top-gated 'fractional region'. The filling of the 'fractional region' is varied by applying negative bias to the gate (colored in dark green). The current flows along a mesa 'defining gate' (dark blue) negatively biased. The transmitted current is deduced from the measured voltage between P1 and the ground. Each sample embeds two 'fractional regions', a long region (LR) of 45µm and a short region (SR) of either 4µm or 0.4µm.

a

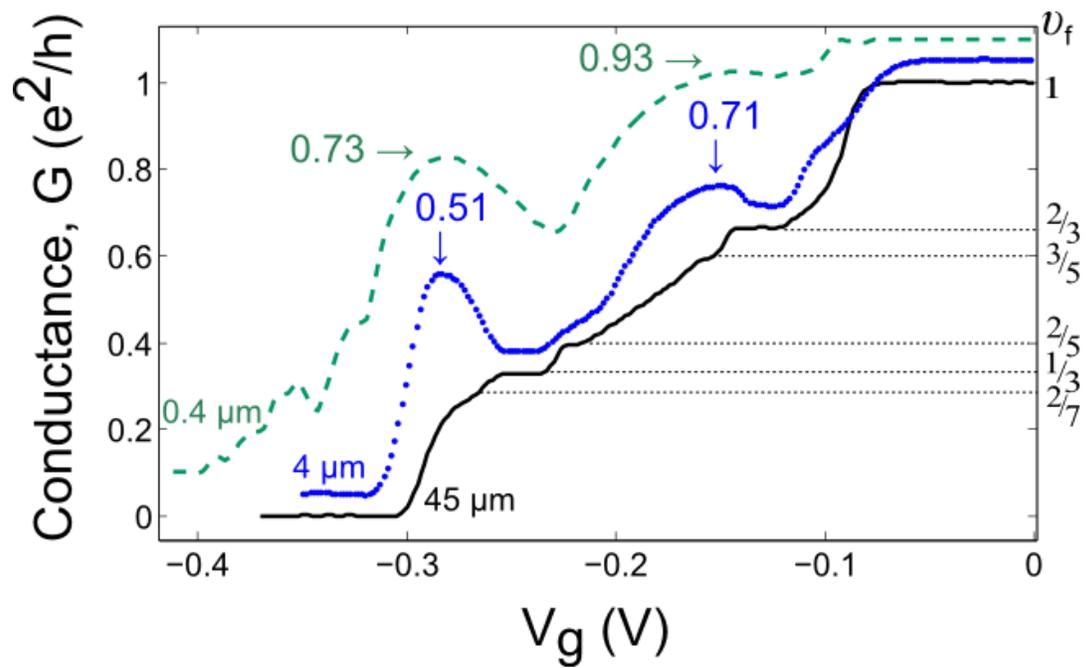

b

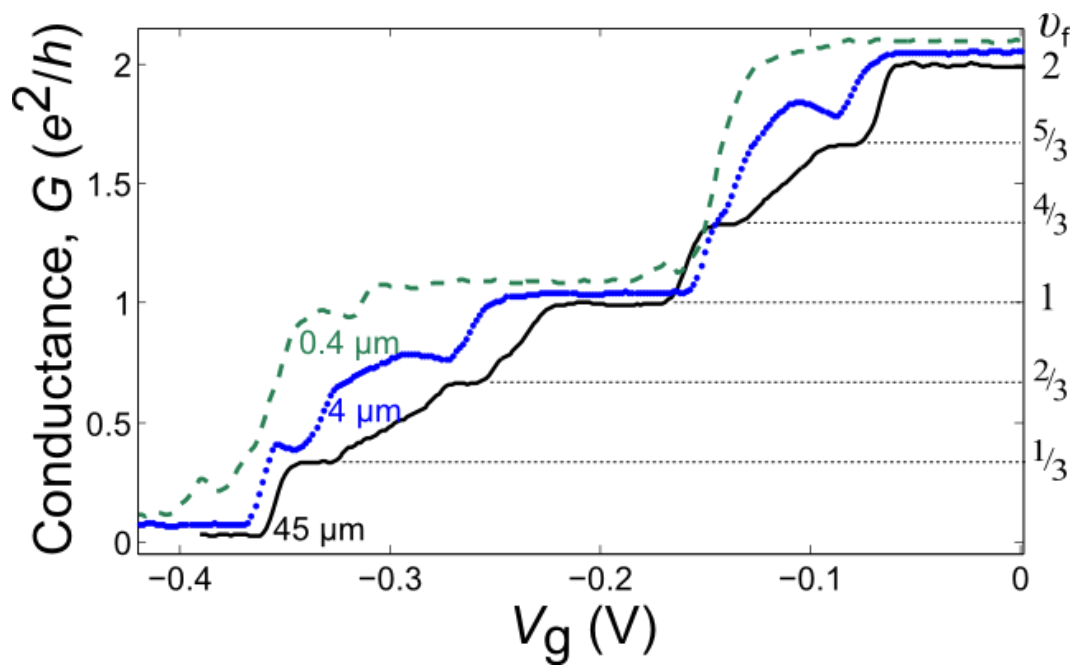

**FIG. 2**. Conductance vs. 'Top Gate' voltage of three different 'fractional regions' sizes: Long region (LR) - 45μm (black solid), short region (SR) - 4μm (blue dotted), short region (SR) - 0.4μm (dark cyan dashed). a) Measurements performed at 8.5T and bulk filling factor $v_B$=1. The LR conductance shows a few fractional plateaus. For SR=4μm the conductance near $v_f$=2/3 rises up to $0.71 \times e^2/h$, and of SR=0.4μm peaks at $0.93 \times e^2/h$. Similarly, the conductance near $v_f$=1/3 peaks at $0.51 \times e^2/h$ for SR=4μm and $0.73 \times e^2/h$ for SR=0.4μm. Conductance plateaus at $v_f$=2/5, 3/5 and a signature of $v_f$=2/7 also seen at LR. For $v_f$=2/5 a reminiscent signature of reconstruction is also seen in the SR, while at $v_f$=3/5, any signature is absorbed by the conductance peak of $v_f$=2/3. The graphs are shifted by $0.05 \times e^2/h$ for convenience. b) Similar measurements at 4.3T at $v_B$=2. Conductance of LR (black solid line) shows plateaus at values of (2, 5/3, 4/3, 1, 2/3, 1/3)$\times e^2/h$. The SR=4 μm shows weaker conductance peaks near (5/3, 2/3, 1/3) while around the integer value of $1 \times e^2/h$ the conductance is flat or monotonic, without peaks. For the SR=0.4μm the conductance increases near all the fractional fillings but remains constant $1 \times e^2/h$ near $v_f$=1.

a

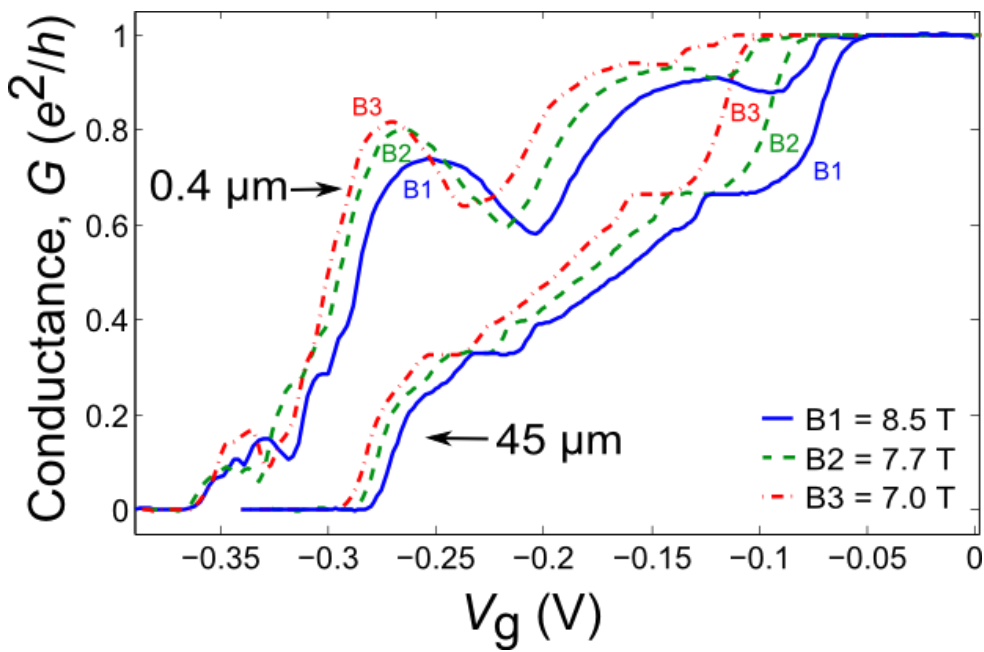

b

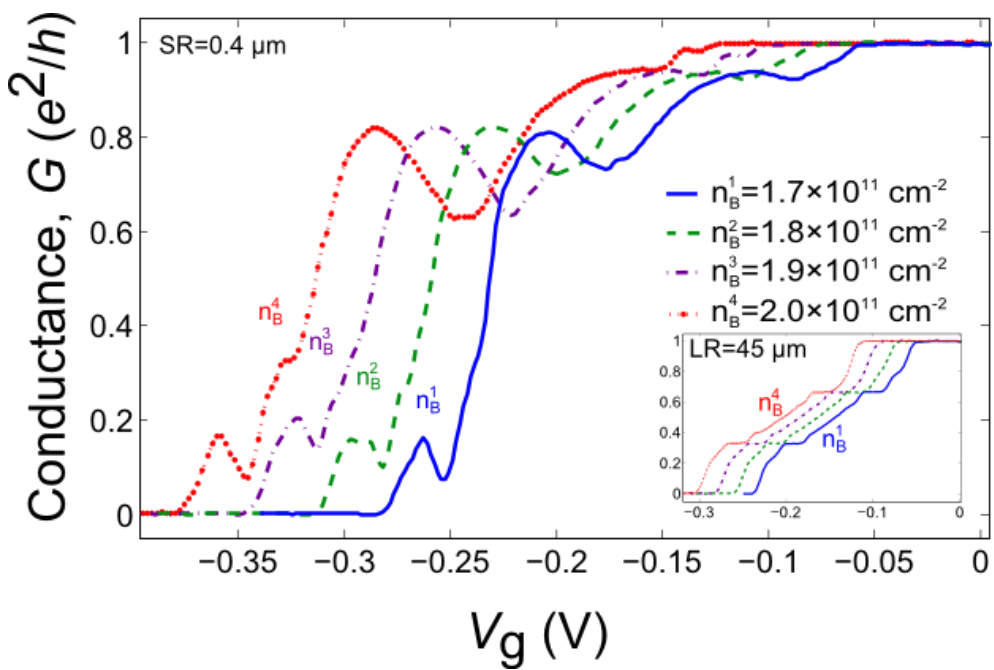

**FIG. 3.** Dependence of conductance on magnetic field and electron density. a) Conductance of LR=45μm and of SR=0.4μm as function of the 'top-gate' voltage at different magnetic fields $B$= (8.5, 7.7, 7.0) T, drawn in blue (solid), green (dash), red (dash-dot), respectively. As the magnetic field decreases the conductance near $v_f$=1/3, 2/3 rises slightly. Similar behavior was observed for SR=4μm (supp.). b) Conductance of LR=45μm (inset) and of SR=0.4 μm as function of the 'top-gate' voltage at $B$=7T and at different bulk densities of $n_B$= (1.7, 1.8, 1.9, 2)×$10^{11}$cm$^{-2}$, corresponding to different bias of the 'back-gate' (0.4, 0.6, 0.8, 1.0) V, drawn in blue (solid), green (dash), purple (dash-dot) and red (dotted), respectively. The conductance peaks for $v_f$=1/3, 2/3 remain nearly unchanged.

a

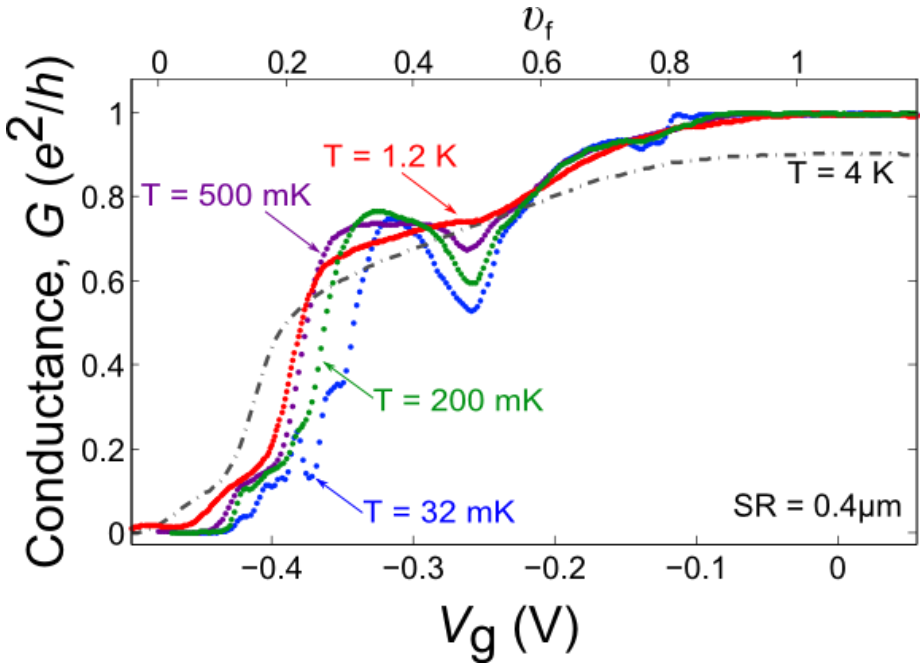

b

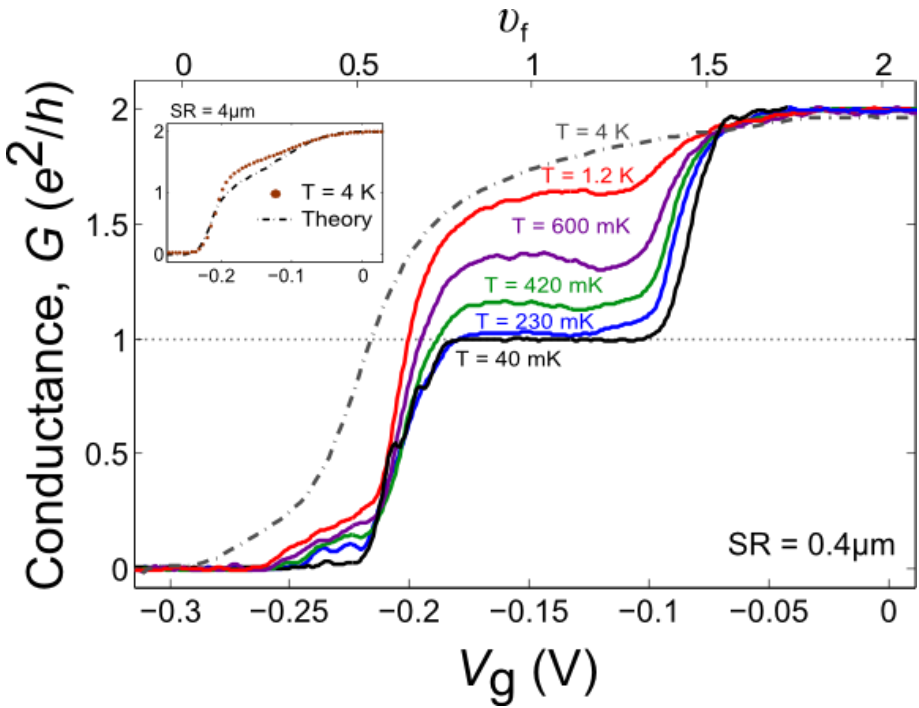

**FIG. 4.** Temperature dependence of SR=0.4µm, 30mK – 4.2K at $v_B$=1 and $v_B$=2. The bottom *x* axis denotes the voltage on the top gate, the top *x* axis is the corresponding filling factor $v_f$ at SR. a) Conductance of SR=0.4µm as a function of 'top gate' voltage at different temperatures (32 mK, 200 mK, 500 mK, 1.2K, 4.2K). The apparent conductance peaks near $v_f$=1/3, 2/3 decay as temperature rises, with that of the $v_f$=1/3 being much more prominent. b) Similar conductance measurements at $v_B$=2 with temperature range of 40mK, 420mK, 600mK, 1.2K, 4.2K. As temperature increases throughout the sample, the transmitted current in $v_f$=1 rises due to increase of electron tunneling through the gated regions, as expected. Inset: Comparison of the classical simulation and the high temperature measurement at SR=4µm; - being qualitatively similar.

# Non-Equilibrated Counter Propagating Edge Modes in the Fractional Quantum Hall Regime – Supplementary Information

Anna Grivnin, Hiroyuki Inoue, Yuval Ronen, Yuval Baum, Moty Heiblum, Vladimir Umansky and Diana Mahalu

## I. Additional 'fractional region' sizes

Conductance vs. top gate voltage as measured at base temperature, $B=8.5T$. The conductance rises non-monotonically at various sizes of the gated region, confirming this effect does not occur due to geometrical current resonance.

a

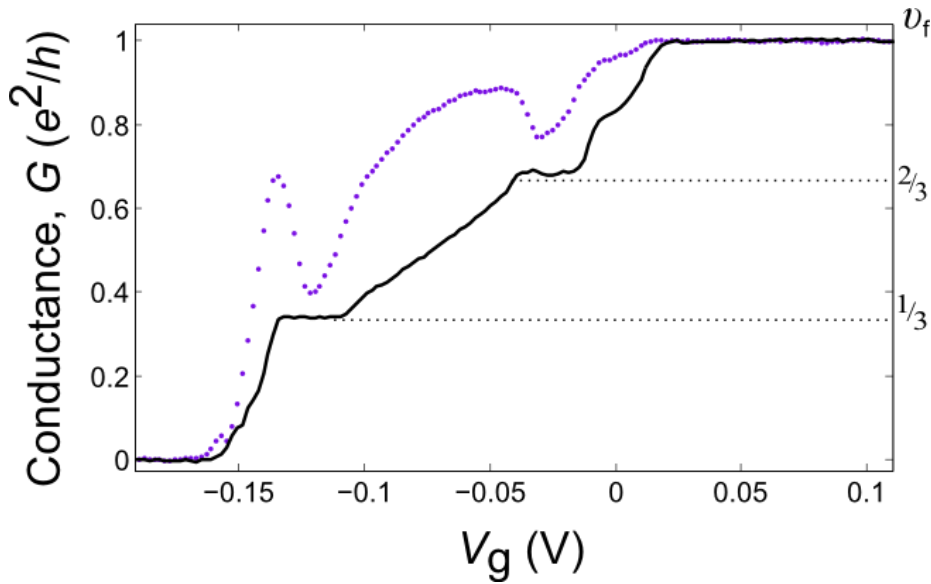

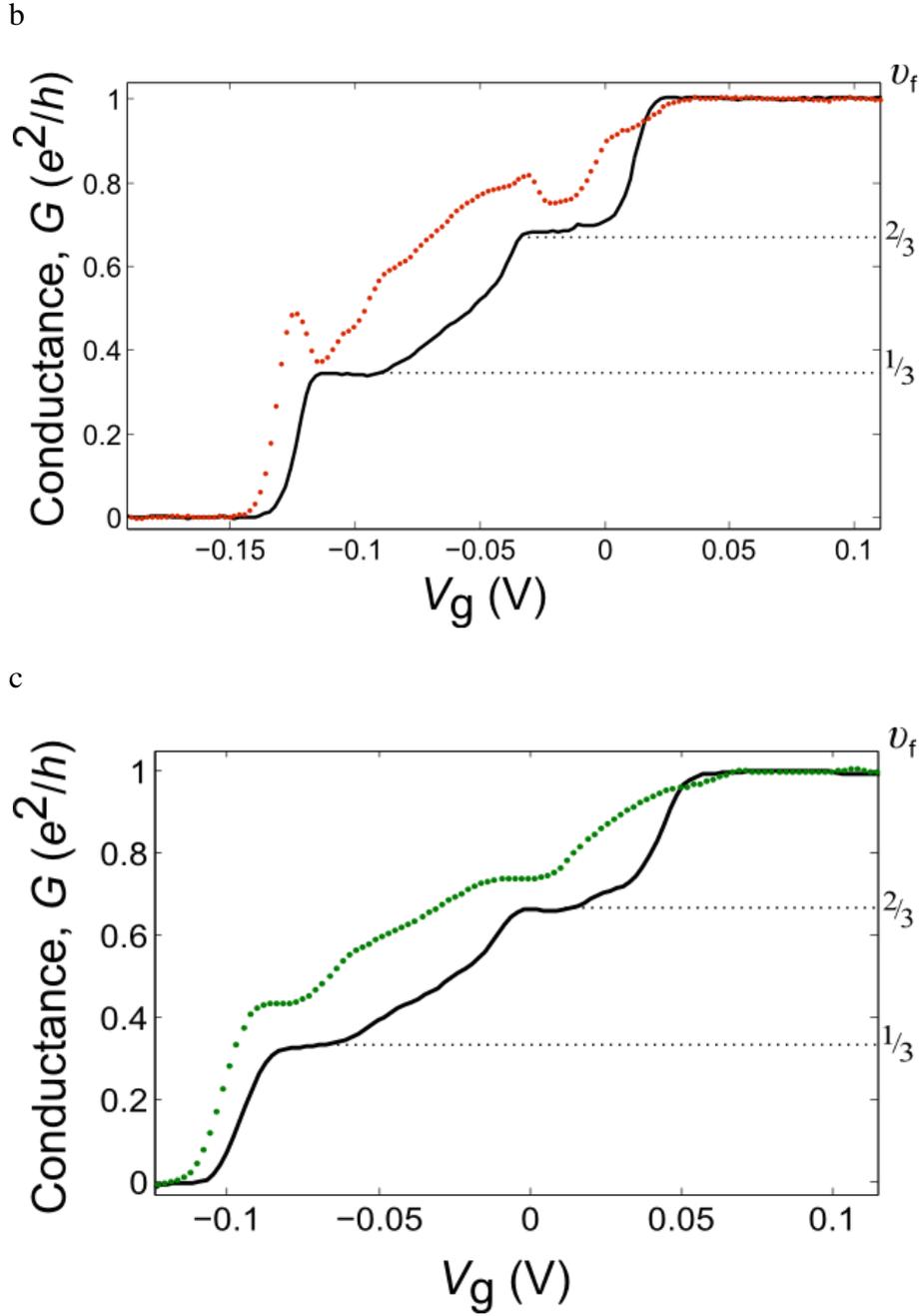

**FIG. S1.** Conductance measurements through additional sizes of the 'fractional region' not studied in the main text of 1, 3, and 7μm, shown in figures a (purple dotted), b (orange dotted) and c (green dotted) correspondingly. In all figures the conductance is compared to conductance of the LR (45μm). We observe conductance rise, which is the strongest for the 1μm SR and weakens as the 'fractional region' is wider. For the width of 7μm the reconstruction decays completely. The measurements were done at magnetic field of 8.5T and density of $1.9 \times 10^{11} cm^{-2}$.

## II. Integer filling factors

a

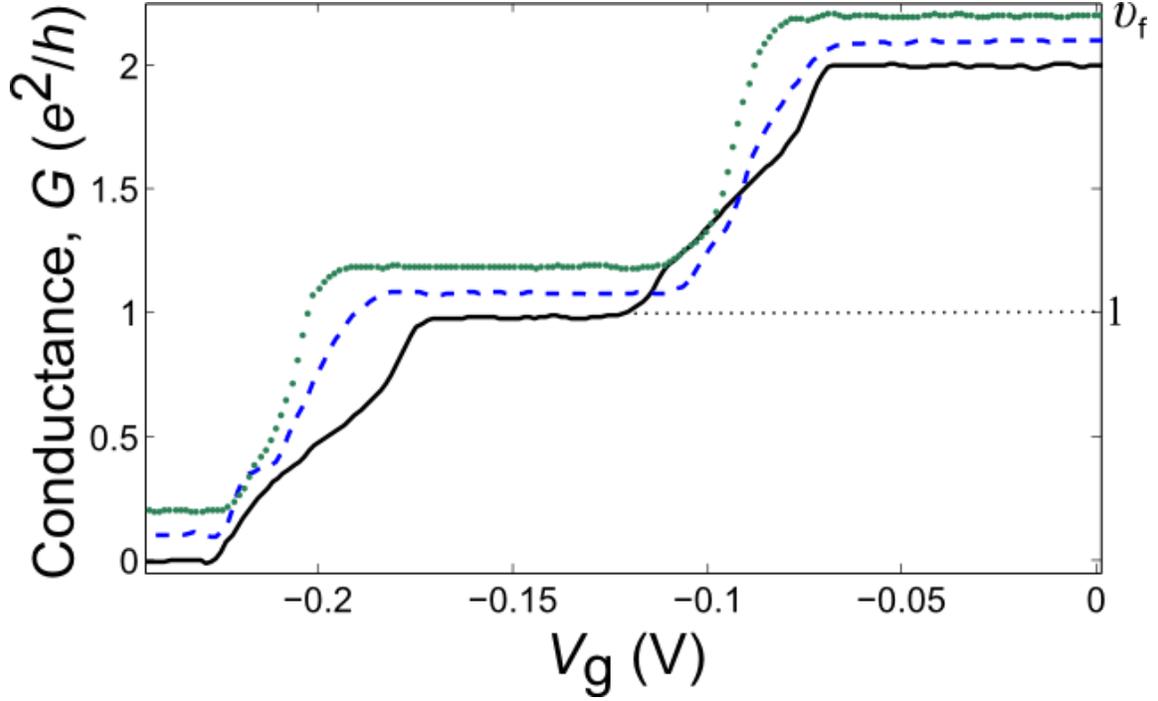

**FIG. S2.** No reconstruction effect is seen in integer fillings of the gated region. Conductance through LR (black solid) and both SR's (4µm and 0.4µm of length, blue and dark cyan respectively) vs. gate voltage measured continuously from $v_f = 2$ to zero. The figure shows conductance as was measured at $B=2.8$T at base temperature $n_B=1.4 \times 10^{11}$cm$^{-2}$. The conductance shows conductance steps at all three sizes of the gated regions at $2 \times e^2/h$ and $1 \times e^2/h$. The graphs are shifted by $0.1 \times e^2/h$ for the reader's convenience.

## III. Magnetic Field and Density Dependence of 4μm 'fractional region'

a

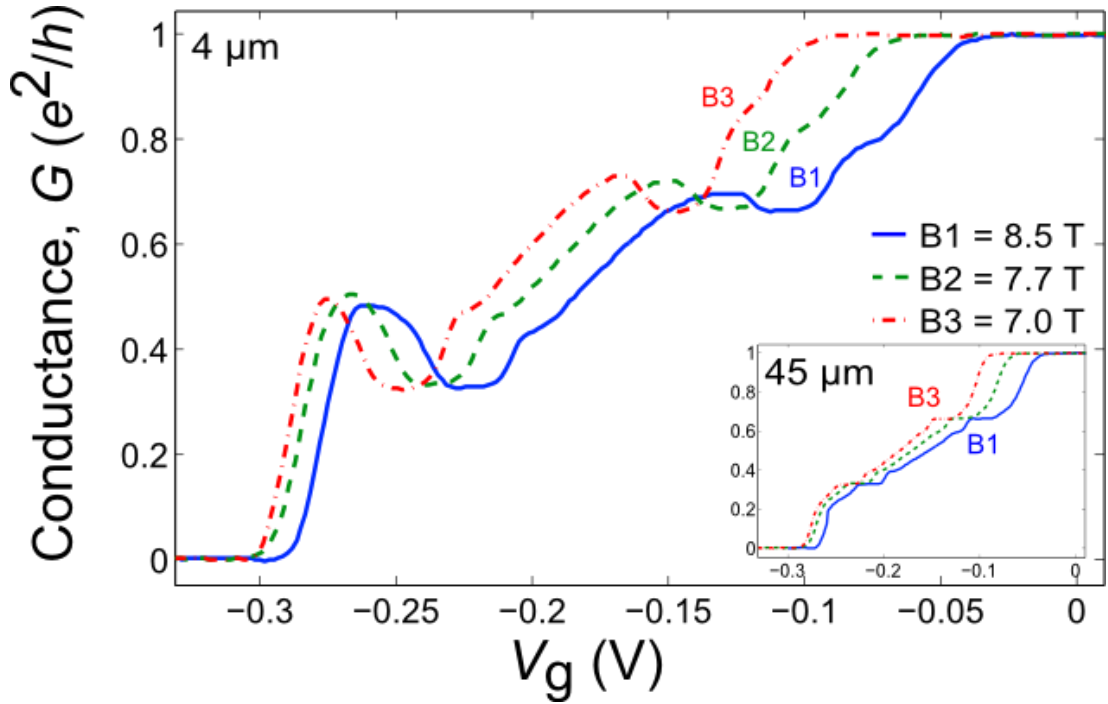

b

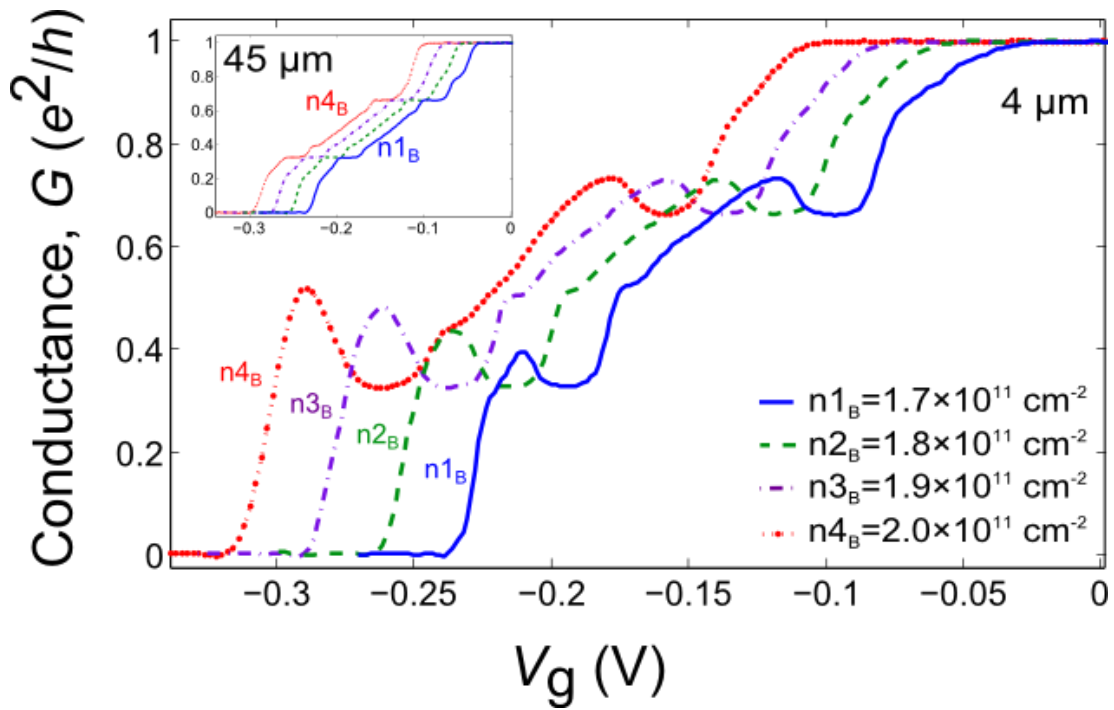

**FIG. S3.** Conductance dependence on magnetic field and density of 4μm 'fractional region'.

a) Conductance of LR - 45μm (inset) and of SR=4μm as function of the top-gate voltage at different magnetic fields (8.5, 7.7, 7.0) T drawn in blue (solid), green (dash), red (dash-dot) correspondingly. As the magnetic field decreases the conductance near $v_f=2/3$ rises slightly while the conductance near $v_f=1/3$ does not change significantly.

b) Conductance of LR (inset) and of SR=4μm as function of the top gate voltage at different bulk densities of $(1.7, 1.8, 1.9, 2) \times 10^{11} \text{cm}^{-2}$ corresponding to different bias on the back-gate (0.4, 0.6, 0.8, 1.0) V drawn in blue (solid), green (dash), purple (dash-dot) and red (dotted) correspondingly, as was measured at 7T. The conductance peaks corresponding to $v_f=2/3$ remain unchanged as density decreases while the conductance peaks near $v_f=1/3$ decay with bulk density.

## IV. Classical Model

We consider transport in a two-dimensional sample with the following simplified geometry:

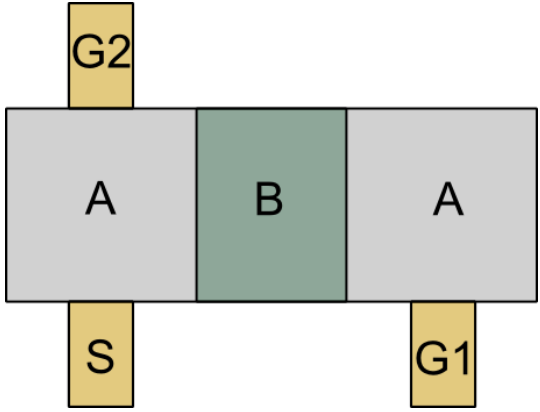

The yellow square regions imitate the ohmic contacts, which we constrain to be at constant electric potential, where contact G1 and G2 are grounds hence at zero potential and contact S is the source and is at constant potential which we choose to be unity for simplicity.

We define in region A the conductivity matrix to be $e^2/h \times \begin{pmatrix} 0 & 2 \\ -2 & 0 \end{pmatrix}$.

In region B we define the conductivity matrix $e^2/h \times \begin{pmatrix} \sigma_{XX} & 2 \cdot v \\ -2 \cdot v & \sigma_{XX} \end{pmatrix}$ where $\sigma_{XX} \ll 1$.

Ideally, $\sigma_{XX} = 0$ its finite value is a numerical constrain which cannot be avoided; However, it does not affect the final results. $v$ is the filling factor at region B which is varied by changing the density in B, from 1 to zero. At the contacts regions there is no Hall resistance and $\sigma_{XX} \gg 1$.

We solve numerically the classical Poisson equation $\nabla(\sigma \cdot \nabla\varphi) = 0$ for the electric potential $\varphi$, imposing boundary conditions: No current is allowed to flow in the perpendicular direction to sample walls, the potential at G1 and G2 must be zero. We find the total current transmitted as function of the filling factor at region B. We translate the dependence on the filling factor to gate voltage dependence using the measurements of the conductance vs. Vg of the wide 'fractional region'.

The potential distribution of the sample for a wide region B (corresponding to the 45μm 'fractional region') is presented below, followed by the potential distribution for a short region B (corresponding to the 4μm 'fractional region'). The current, correspondingly, flows along equipotential lines.

a

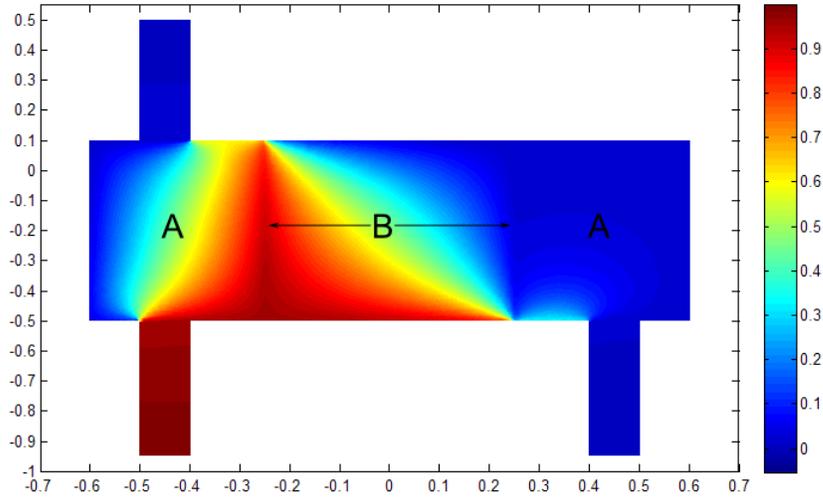

b

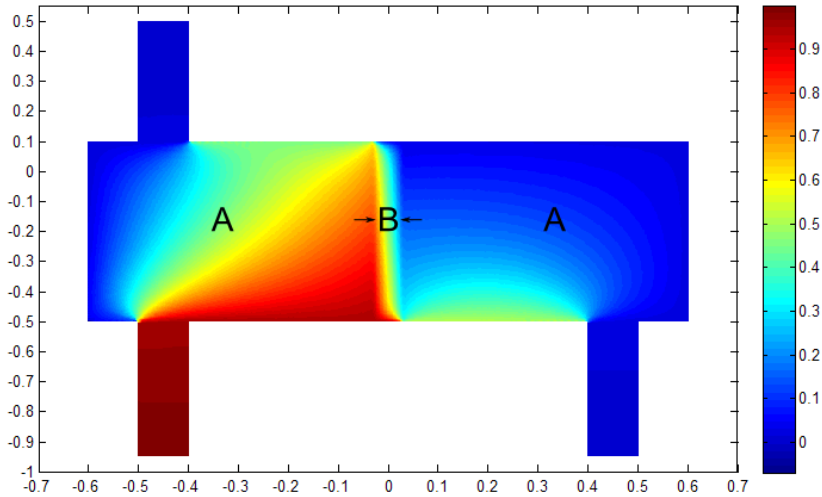

**FIG. S4.** Voltage potential at the sample region as calculated from the classical equations solution. a) For a wide region B (corresponding to the 45μm 'fractional region') b) For a short region B (corresponding to the 4μm 'fractional region').

# V. Temperature dependence – $v_B=2$

Here we present the measured conductance across the short region (SR = 4μm) at bulk filling $v_B=2$ ($B = 3.5$T) as function of temperature, and compare to the classical solution of the Poisson equation. As temperature rise the conductance approaches the classically predicted conductance dependence on the top gate voltage. We also observe that at certain temperatures around 750mK the conductance is non-monotonic due to significant bulk conductance contributing to the total current.

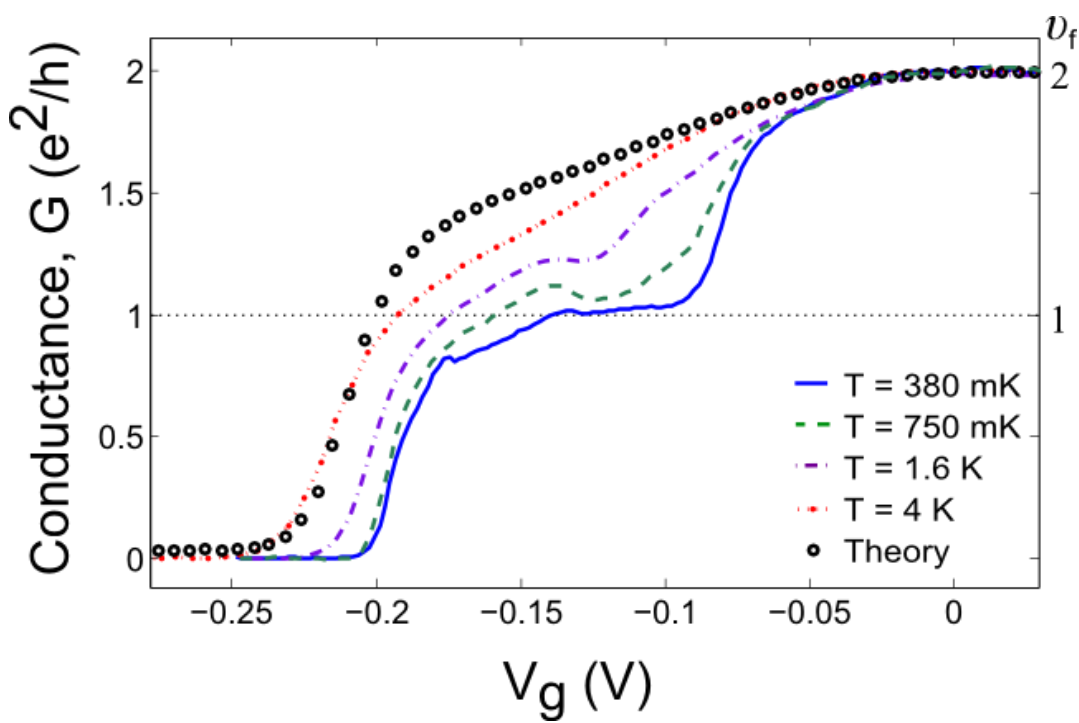

**FIG. S5.** The temperature dependence of the conductance across SR=4μm measured at 380mK, 750mK, 1.6K, and 4K drawn in blue (solid) green (dashed) purple (dash-dot) and red (dotted) correspondingly, the theoretical prediction is plotted in empty black circles.

## VI. Temperature dependence - $v_B=1$

Temperature dependence of the conductance across 4μm 'fractional region' and comparison to the theoretical curve (at filling factors close to zero under the 'fractional region' the model breaks down so that it's not possible to estimate the conductance there accurately). The bulk filling is $v_B=1$ and $B=8.5T$.

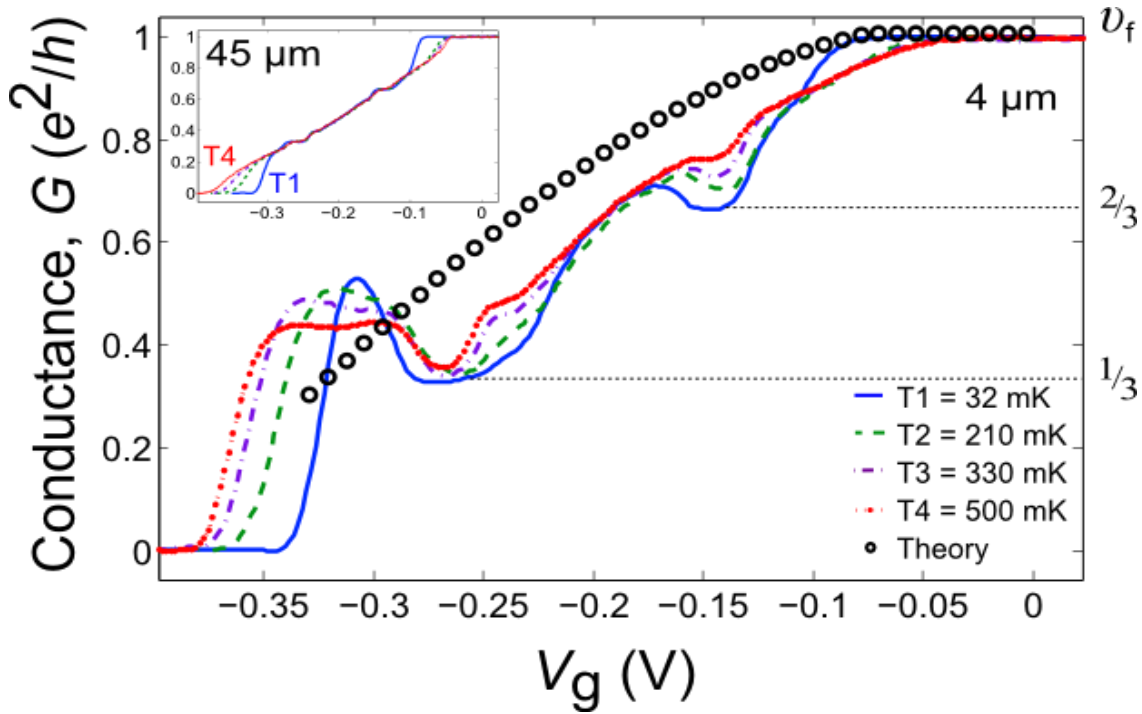

**FIG. S6.** The temperature dependence of the conductance across SR=4μm measured at 32mK, 210mK, 330mK, and 500mK drawn in blue (solid) green (dashed) purple (dash-dot) and red (dotted) correspondingly, the theoretical prediction is plotted in empty black circles. Here, the conductance at low temperatures overshoot the classical prediction, indicating there is a reconstruction effect contributing to the current rise rather than classical bulk contribution, which we observed for the integer case.

## VII. Bulk filling factor of 2/3

We tuned the back-gate voltage and the magnetic field to $v_B=2/3$ and scanned the top gates varying $v_f$ continuously from $v_f=2/3$ to zero. The transmission under LR shows a quantized plateau at $v_f=1/3$. The conductance through both SR's shows enhancement at the region of $v_f=1/3$ similarly to the $v_B=1$ case. The conductance through SR=4μm rises up to $G = 0.42 \times \frac{e^2}{h}$ and the conductance of SR=0.4μm peaks at $G = 0.59 \times \frac{e^2}{h}$. Consequently, we conclude the effect of conductance increase occurs whenever the $v_f$ is at fractional Hall state without dependence on the bulk filling factor – whether it is integer or fractional.

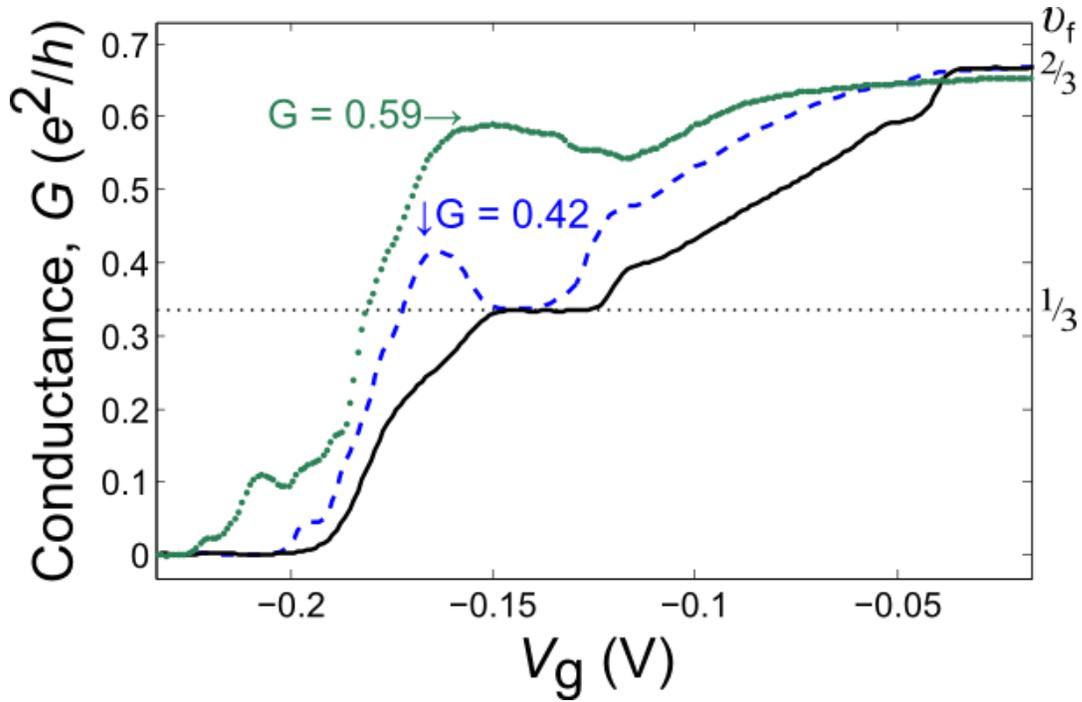

**FIG. S7.** Conductance through LR (black) and both SR's (4μm in blue dashed and 0.4μm in dark cyan dotted) vs. gate voltage measured continuously from $v_f=2/3$ to zero. The sample itself is at $v_B=2/3$. The conductance shows peak near $v_f=1/3$ that rises up to $0.42 \times \frac{e^2}{h}$ for SR=4μm and to $0.59 \times \frac{e^2}{h}$ for SR=0.4μm.